\documentclass[11pt,twoside]{article}
\usepackage{hepro5}
\usepackage{graphicx}

\usepackage[T1]{fontenc} 

\usepackage{latexsym}
\usepackage{verbatim}

\usepackage{amssymb,amsmath}
\usepackage{natbib}
\begin{document}

\vskip 1.0cm
\markboth{G.~Morlino et al.}{Mass-loading of bow shock PWN}
\pagestyle{myheadings}

\vspace*{0.5cm}
\title{Mass-loading of bow shock pulsar wind nebulae}

\author{G.~Morlino$^1$, M.~Lyutikov$^2$ and M.~Vorster$^2$}
\affil{$^1$INFN -- Gran Sasso Science Institute, viale F. Crispi 7, 67100 L'Aquila, Italy\\
$^2$Department of Physics, Purdue University, 525 Northwestern Avenue, West Lafayette, IN 47907-2036, USA}

\begin{abstract}
We investigate  the dynamics of bow shock nebulae created by pulsars moving supersonically through a partially ionized interstellar medium.  A fraction of interstellar neutral hydrogen atoms penetrating into the tail region of a pulsar wind will undergo photo-ionization due to the UV light emitted by the nebula, with the resulting mass loading dramatically changing the flow dynamics of the light leptonic pulsar wind.  Using a quasi 1-D hydrodynamic model of relativistic flow we find that if a relatively small density of neutral hydrogen, as low as $10^{-4}$ cm$^{-3}$, penetrate inside the pulsar wind, this is sufficient to strongly affect the tail flow.  Mass loading leads to the fast expansion of the pulsar wind tail, making the tail flow intrinsically non-stationary.  The shapes predicted for the bow shock nebulae compare well with observations, both in H$\alpha$ and X-rays.  
\end{abstract}

\section{Introduction}
 \label{intro} 

It has been estimated that between $10\%$ and $50\%$ of pulsars are born with kick velocities $V_{\rm NS} \gtrsim 500$ km s$^{-1}$. These pulsars will escape from their associated supernova remnants into the cooler, external  interstellar medium (ISM) in less than 20 kyr  \citep{Cordes98,Arzoumanian2002}.  As this time scale is sufficiently short, the pulsars are still capable of producing  powerful relativistic winds.  Furthermore, comparison with typical sound speeds in the ISM, $c_{s, \rm ISM} = 10-100$ km s$^{-1}$, shows that the pulsars are moving with highly supersonic velocities.  The interaction of the pulsar's wind with the ISM produces a bow shock nebula with an extended tail. If a pulsar is moving through a partially ionized medium, the bow shock nebula can be detected by the characteristic H$\alpha$ emission resulting from the collisional  and/or charge-exchange  excitation of neutral hydrogen atoms in the post-shock flows and the subsequent emission via bound-bound transitions \citep{Chevalier80}.  To date, nine such bow shock nebulae emitting H$\alpha$ have been discovered, including three around $\gamma$-ray pulsars \citep{Brownsberger14}. 

\begin{figure}
\begin{center}
\includegraphics[width=0.49\textwidth]{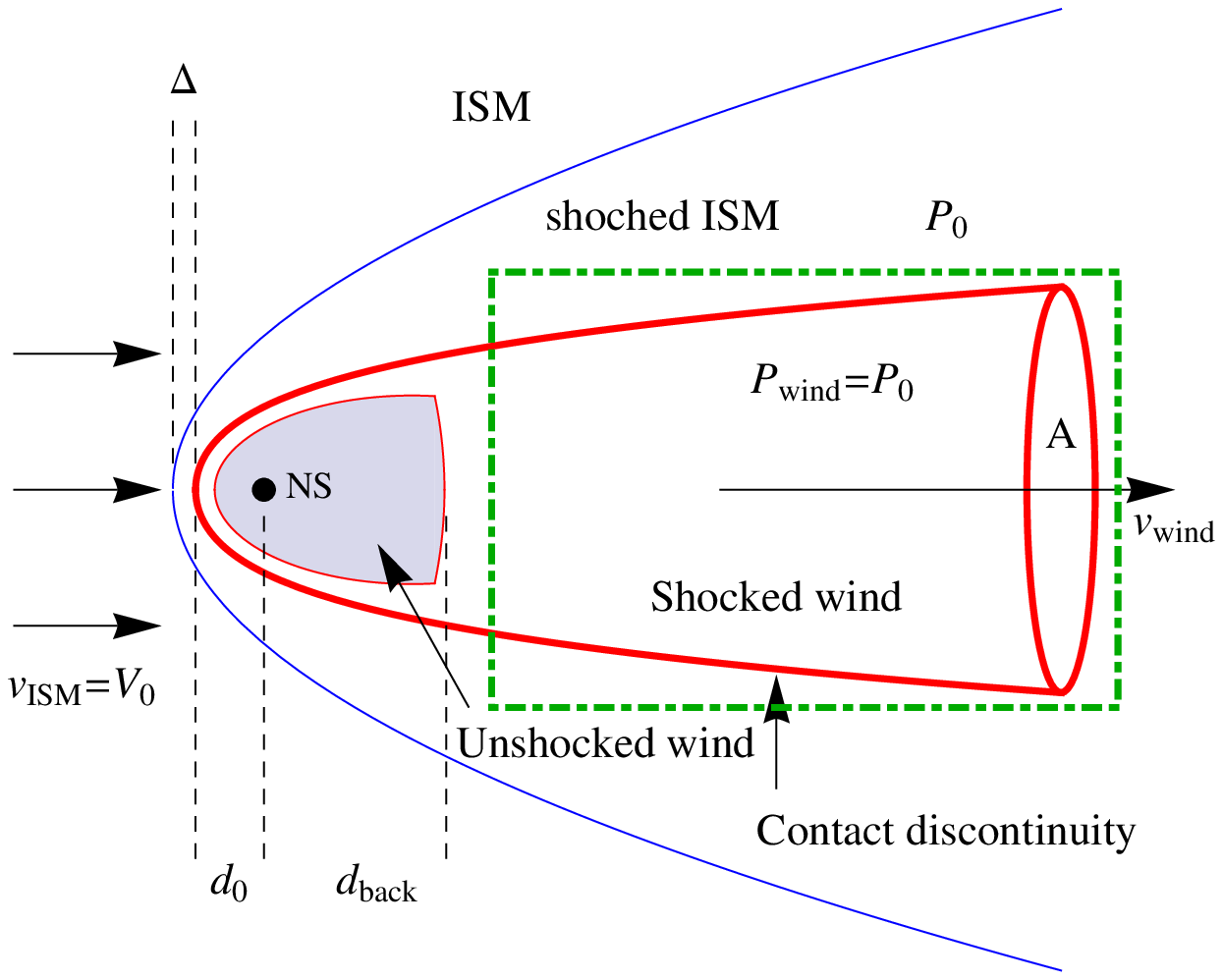}
\includegraphics[width=0.49\textwidth]{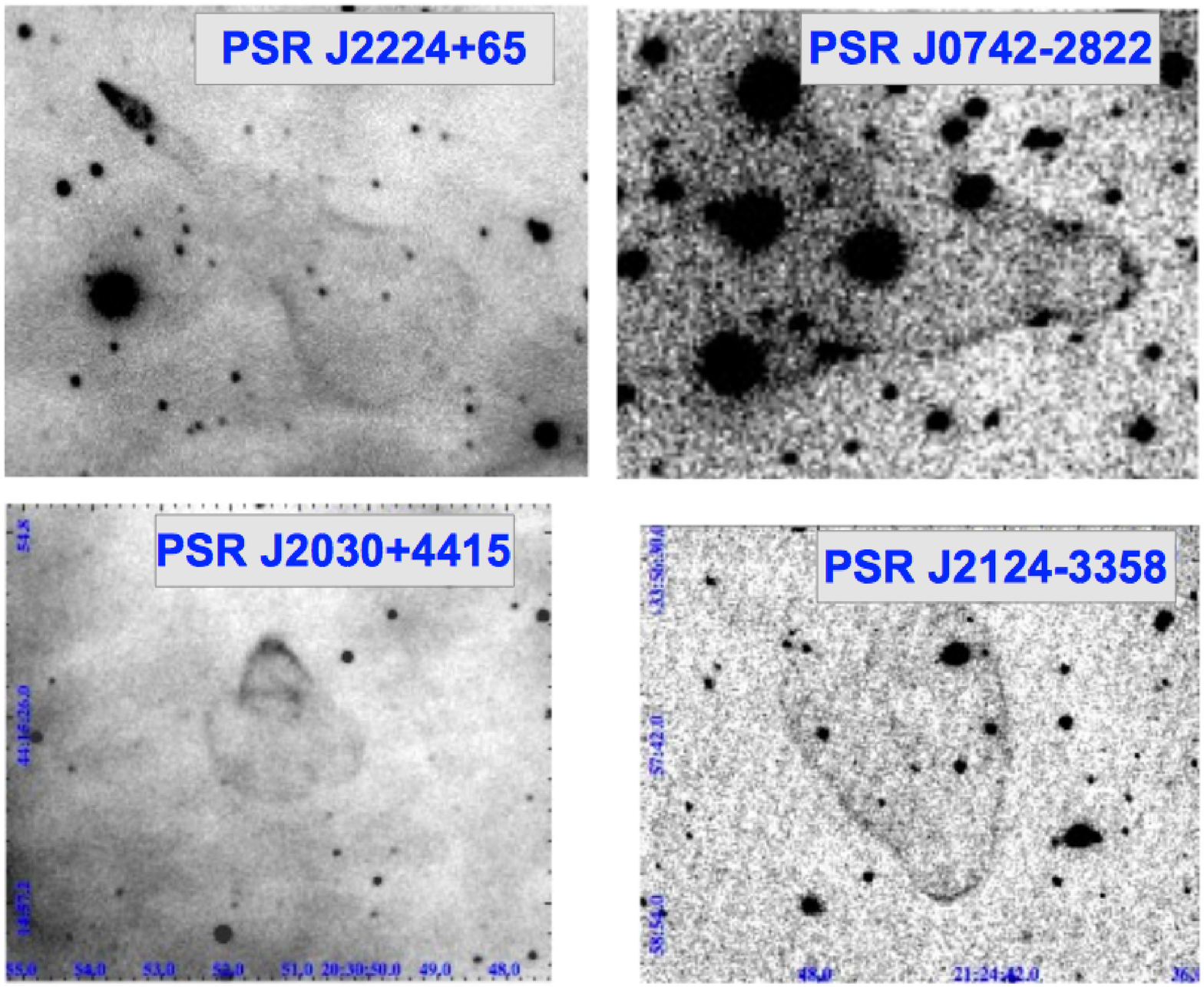}
\end{center}
\caption{{\it Left}: schematic illustration of a bow shock nebula propagating through fully ionized ISM, as seen in the rest frame of the pulsar. The dot-dashed box identify the region studied in \S\ref{sec:structure}. {\it Right}: montage of H$\alpha$ images of optical bow shocks associated with pulsar wind nebulae. Shown are J2224+65, the so-called {\it Guitar nebula} \citep{Chatterjee02}, J0742-2822, J2030+4415, and J2124-3348 \citep{Brownsberger14}.}
\label{fig1}
\end{figure}

Hydrodynamic (and hydromagnetic) models of bow shock nebulae predict the formation of a smooth two shock structure \citep{Bucciantini02a} schematically shown in Fig.\ref{fig1} (left panel): a forward shock in the ISM separated by a contact discontinuity from a termination shock in the pulsar wind.  In the head of the nebula the contact discontinuity is situated at a distance given by Eq.(\ref{eq:d0}), corresponding to the position where the ram pressure of the ISM balances the pulsar wind pressure.  The flow structure in the head of the nebula is  reasonably well understood, especially in the limit of strong shocks \citep{Wilkin96}.

In contrast to these numerical models, H$\alpha$, radio and X-rays observations show that the morphologies of bow shock nebulae are significantly more complicated.  More specifically, observations reveal that the tails of bow shock nebulae have a highly irregular morphology: Fig.\ref{fig1} (right panel) shows four such examples.  All these nebulae have a characteristic \emph{``head-and-shoulder''} structure, with the smooth bow shock in the head not evolving into a quasi-conical or  quasi-cylindrical shape, but instead showing a sudden sideways expansion(s).  Arguably the most famous example is the {\it Guitar nebula} powered by the pulsar PSR B2224+65 (see right panel in Fig.\ref{fig1}).  As the name suggests, this nebula has a guitar-like shape with a bright head, a faint neck, and a body consisting of several larger bubbles.   

These peculiar tail shapes have been interpreted as the result of density variations in the ISM \citep{Romani97,Vigelius07}.  However, several observations {\bf seem} to suggest that the peculiar morphological features could result from the internal dynamics of the pulsar wind, rather than through inhomogeneities in the ISM.  First of all, all tails show {\it similar} morphological variations. Moreover a common characteristic of these bow shock nebulae is that they are all highly symmetric with respect to the direction of motion of the pulsar - this is not expected in general if variations are due to the external medium. Finally, in many cases morphological features in H$\alpha$, radio and X-rays are quasi-periodic. 
It has also been proposed  that the morphology of the Guitar nebula could be explained by (unidentified) instabilities in the jet-like flow of pulsar material away from the bow shock \cite{Kerkwijk08}, even if the magnetic field tension could stabilize the pulsar wind tale \citep{Bucciantini05}.

In the present paper we illustrate the effect of neutral hydrogen on the tail region of these nebulae and we suggest that the mass loading of neutral hydrogen in the pulsar wind can explain the peculiar morphology observed at H$\alpha$, radio and X-ray energies. The detailed analysis of this hypothesis has been presented in \cite{Morlino15}.
In order to focus on the effect of mass loading on the evolution of bow shock nebulae, complications introduced by magnetic field pressure (and topology) are neglected in the present paper.

\section{Interaction of neutrals}
 \label{sec:structure} 
In order for mass loading to play a role in the dynamic evolution of bow shock nebulae, neutrals are required to cross the distance between the bow shock and the contact discontinuity, $\Delta$, and penetrate into the wind region (see Fig.\ref{fig1}). Hence the interaction length inside the shocked ISM must be larger than $\Delta$, which has been estimated to be $\Delta= 5/16 \,d_0$ \citep{Chen96,Bucciantini05}, where $d_0$ is the distance between the pulsar and the contact discontinuity (CD) (formed between the shocked ISM and the shocked wind). The distance $d_0$ is obtained by equating the wind pressure with the bulk pressure of the ISM:
\begin{equation} \label{eq:d0}
 d_0= \left( \frac{\mathcal{L}_{w}}{4 \pi V_{\rm NS}^2  \rho_{\rm ISM} c}  \right)^{1/2}
       = 1.3 \times 10^{16} \mathcal{L}_{w,34}^{1/2} \, V_{300}^{-1} \, n_{\rm ISM,-1}^{-1/2} \, {\rm cm} \,,
\end{equation}
where $\mathcal{L}_{w}= 10^{34} \mathcal{L}_{w,34}\,\text{erg}\,\text{s}^{-1}$ is the pulsar wind luminosity, $V_{\rm NS}= 300\, V_{300}\,\text{km}\,\text{s}^{-1}$ is its peculiar velocity, and $\rho_{\rm ISM}=m_p n_{\rm ISM}$ is the density of the dragged component of the ambient medium, expressed in units of $n_{\rm ISM}= 0.1\, n_{\rm ISM,-1}\,\text{cm}^{-3}$.

After crossing the bow shock, neutral atoms can interact with the shocked protons. The interaction length is given by 
\begin{equation} \label{eq:L_coll}
 \lambda= \frac{V_{\rm NS}}{ X_{\rm ion} n_{\rm ISM} r_c \langle \sigma v_{\rm rel} \rangle} \,,
\end{equation}
where $X_{\rm ion}$ is the ionization fraction of the ISM, $r_c$ is the compression ratio of the bow shock, $\sigma$ is the relevant cross section of the process under consideration, and $\langle \sigma v_{\rm rel} \rangle$ is the collision rate averaged over the ion distribution function. Using the fiducial values $n_{\rm ISM}=0.1\,\text{cm}^{-3}$ and $V_{\rm NS}= 300\,\text{km}\,\text{s}^{-1}$, together with an ionization fraction of 90\% and $r_c=4$ (the typical value for strong shocks), leads to the following estimates for the mean free paths \citep{Morlino15}:
 $\lambda_{{\rm ion},p}  \approx 3.0 \times 10^{20} \, \rm cm   \label{eq:L_ion_p}$,
 $\lambda_{{\rm ion},e} \approx 2.2 \times 10^{16} \, \rm cm   \label{eq:L_ion_e}$, and
 $\lambda_{\rm CE}  \approx 1.5 \times 10^{15} \, \rm cm \,     \label{eq:L_CE}$,
for the ionization due to collisions with protons, electrons and charge-exchange (CE), respectively.  Note that $\lambda_{{\rm ion},e}$ has been calculated under the assumption that the electrons downstream of the shock equilibrate rapidly with protons, thereby acquiring the same temperature.  If this assumption does not hold, the collisional length scale for ionization due to electrons can become much larger. In addition, these values of $\lambda$ are to be taken as lower limits as they are valid just ahead of the nebula, where the compression ratio and the temperature obtain their maximum values. 

From these estimates it follows that only a negligible fraction of the neutral hydrogen will be collisionally ionized by electrons, whereas a significant fraction of neutrals will undergo CE.  The neutrals resulting from a CE event will tend to be dragged with the shocked protons along a direction parallel to the contact discontinuity. Nevertheless, the CE process produces a diffusion of neutrals in the nose of the nebula and it may still be possible for the newly formed neutrals to enter the wind region, provided that their diffusion velocity perpendicular to the contact discontinuity is of the same order or larger than their velocity parallel to the contact discontinuity. This is a complication that will not be addressed in the present paper but is essential to estimate the correct amount of neutrals that can penetrate into the wind. 

Although a large number of neutrals will be lost due to CE, a minimum fraction of neutrals proportional to $\exp[- \Delta/(\lambda_{\rm CE}+\lambda_{\rm ion})]$ will cross the shocked ISM region without suffering any interaction and will enter the pulsar wind in their original state.  These neutrals will not influence the wind structure until they are ionized, either through collisions with relativistic electrons and positrons, or through photo-ionization with photons emitted by the nebula or by the pulsar. In \cite{Morlino15} we showed that photo-ionization due to UV photons emitted by the nebula is the dominant process, while collisional ionization can be neglected.

\section{Hydrodynamical model}
 \label{hydro} 
To study the effect of mass loading  in the tail of bow shock nebulae, we solve, in a quasi 1-D approximation, the steady-state conservation equations for mass, momentum and energy. We restrict our attention to the region enclose in the dot-dashed rectangle of Fig.\ref{fig1}, meaning that any internal structures are neglected, in particular the free wind region and the termination shock.  For the sake of simplicity the presence of the bow shock is also neglected, but we will comment on  possible effects of mass loading on the shape of the bow shock.  
It is further assumed that the external medium has a spatially independent velocity, $V_0$, pressure, $P_0$, and density, $\rho_0$. The quasi 1-D approximation implies that the transverse cross section, $A$, of the flow can change, but that all the characteristic quantities of the wind, \emph{i.e.}, the velocity, $u$, density, $\rho$ and pressure, $P$, are assumed to be a function of the position $x$ only.  

The relativistic equations for the conservation of particle number, energy and momentum, for a 1-D system, written in the rest frame of the neutron star are
\begin{align} 
 \partial_x \left[ n_{e,p} u A \right]                            & = \dot n A' \,,  \label{eq:flux0_n} \\
 \partial_x \left[ w \gamma_w u A \right]                  & = q c^2 \gamma_0 A' \,,   \label{eq:flux0_en} \\
 \partial_x \left[ w u^2 A \right] + c^2 A\partial_x P  & = q c^2 \gamma_0 A' V_0 \,.  \label{eq:flux0_mom}
\end{align}
Here $n_{e,p}$ is the numerical density of electrons (protons), $w$ is the total wind enthalpy and $\gamma_w$ and $\gamma_0$ are the Lorentz factors of the wind and the neutrals, respectively. $A'$ represent the effective area crossed by neutrals. For the sake of clarity we discuss here only the case where the neutrals penetrate everywhere in the nebula, which imply $A'=A$. The mass loading term is given by $q= \dot{n} (m_e+m_p)\approx \dot{n} m_p$, where $\dot{n}$ is the number of hydrogen atoms ionized per unit time and unit volume and is determined by the photo-ionization due to UV photons. Hence $\dot n= n_{N} n_{\rm ph} \bar\sigma_{\rm ph} c$, where $\bar\sigma_{\rm ph}$ is the photo-ionization cross section averaged over the photon distribution whose density is $n_{\rm ph}$, while $n_N$ is the density of neutrals entering inside the wind. Both $n_{\rm ph}$ and $n_N$ are assumed to be constant inside the wind.

The pulsar wind after the termination shock is only marginally relativistic with a bulk Lorentz factor $\gamma_w \approx 1$. Furthermore, the neutrals are non-relativistic, hence $\gamma_0 \approx 1$, while the steady-state assumption requires the pressure to be constant everywhere, \emph{i.e.}, $P=P_0$ and $\partial_x P=0$ (we are assuming a subsonic flow). 

In order to close the system (\ref{eq:flux0_n})-(\ref{eq:flux0_mom}), an expression for the enthalpy is required.  It is generally believed that the shocked pulsar wind predominantly consists of electron-position pairs with highly relativistic temperatures, and we assume that when mass loading occurs there is no energy transfer between electron and protons. In other words, electrons and protons do not reach a thermal equilibrium, but evolve independently with different temperatures. This implies that the electron gas is always highly relativistic, hence the rest mass contribution to enthalpy can be neglected, \emph{i.e.}, $w_e= \epsilon_e+P_e= 4P_e = 4 P_0$. On the other hand, protons are non-relativistic, hence their thermal energy is always negligible with respect to their rest mass energy, and their enthalpy is $w_p = \rho_p c^2$. The total enthalpy of the wind is thus $w = w_p + w_e = \rho_p c^2 + 4 P_0$.

Using all the simplifications outlined above one can obtain an analytical solution for the wind dynamics \citep{Morlino15}. It turns out that the cross section and the proton density can be expressed as a function of the flow speed only:
\begin{align}  
 A(u)          & = u_1 A_1 / u \,, \label{eq:A(u)} \\
 \rho_p(u)  & = 4 P_0/c^2 \, (u_1-u)/(u-V_0) \label{eq:rho(u)} \,,
\end{align} 
where the subscript $_1$ identifies the values of quantities at the initial position $x=x_1$ right behind the backward shock. Notice that, as a consequence of the mass loading, the usual conservation equation $\rho u A = {\it const}$ does not hold any more, being replaced by $u A = {\it const}$.
The flow speed can be expressed as a function of $x$ in an implicit form as follows
\begin{equation} \label{eq:u(x)_relB}
 \frac{x}{\lambda_{\rm ML}} = \frac{u}{u-V_0} - \frac{u_1}{u_1-V_0} + \ln \left[ \frac{u_1-V_0}{u-V_0} \right] \,,
\end{equation}
where we have introduced the mass loading length scale 
\begin{equation} \label{eq:lambda_ML}
 \lambda_{\rm ML} = \frac{4 P_0 (u_1-V_0)}{q c^2} \simeq \frac{4 P_0}{\rho_N c^2} \frac{u_1}{n_{\rm ph} \bar \sigma_{\rm ph} c}  \,.
\end{equation}
The last equality holds because $u_1\approx c \gg V_0$. Eq.(\ref{eq:lambda_ML}) reveals the meanings of $\lambda_{\rm ML}$: this is the typical length scale such that the enthalpy of the loaded mass, $\rho_N c^2$, times $\lambda_{\rm ML}/\lambda_{\rm interaction}$, equals the enthalpy of the wind, $4 P_0$.
Choosing realistic values for the parameters associated with bow shock nebulae emitting H$\alpha$ shows that $\lambda_{\rm ML} \sim 10^{15}$ cm \citep{Morlino15}, even if this value can vary by orders of magnitude, essentially due to the fact that the values of the neutral density inside the wind and the luminosity of the PWN tail are difficult to estimate.

\begin{figure}
\begin{center}
\includegraphics[width=0.49\textwidth]{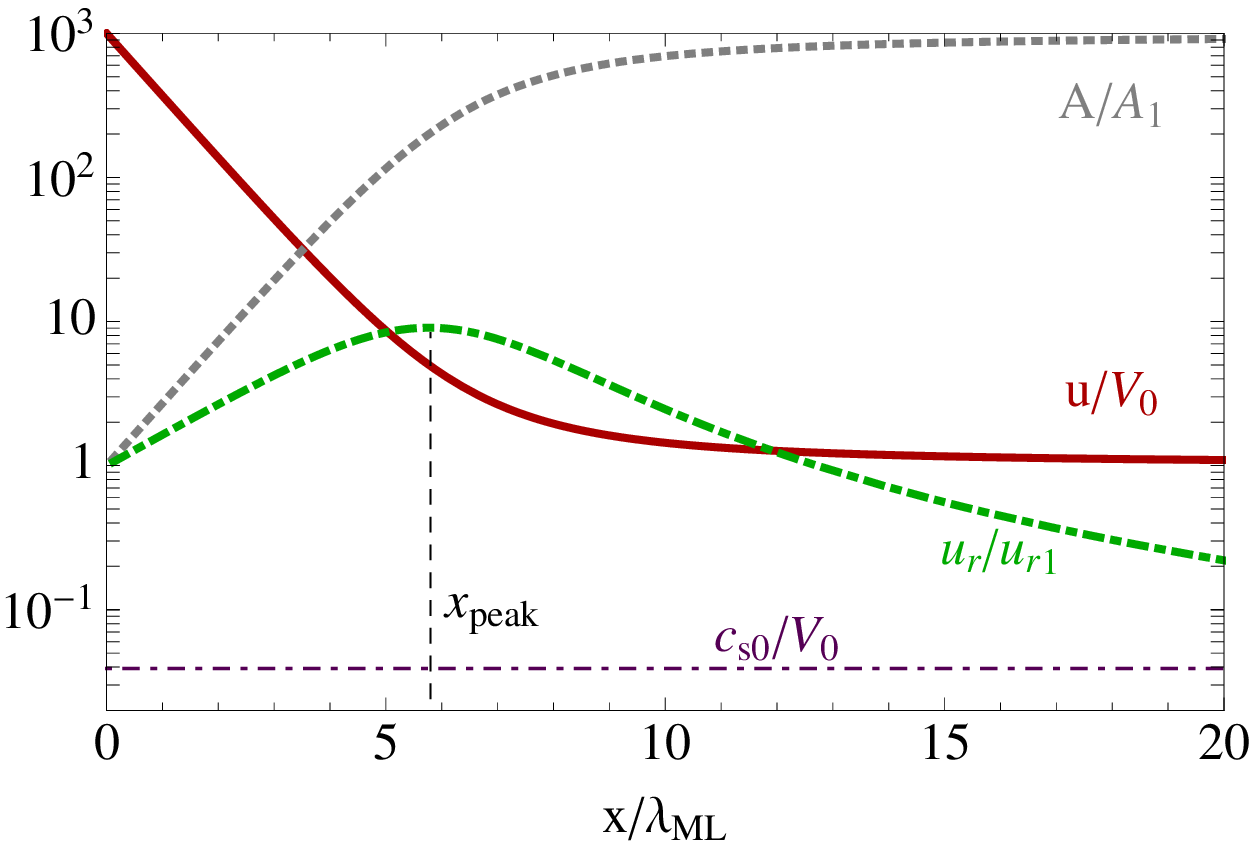}
\includegraphics[width=0.49\textwidth]{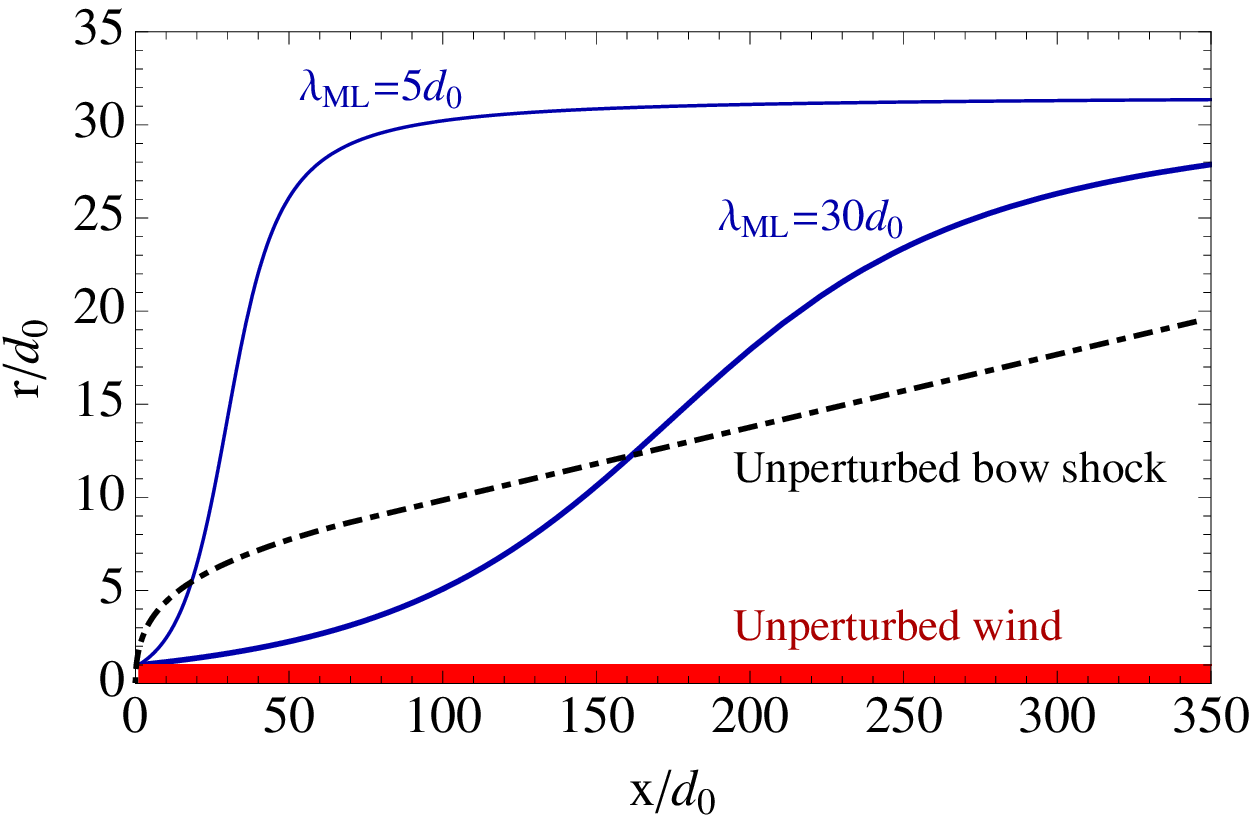}
\end{center}
\caption{{\it Left}: structure of a pulsar wind when mass loading occurs. The various lines represent the wind speed divided by $V_0$ (solid line), wind cross section (dotted line) and radial expansion speed (dot-dashed line) both normalized to their initial values. The sound speed of the ISM, $c_{s0}$, also normalized to $V_0$, is shown for comparison (thin dot-dashed line).
{\it Right}: comparison between two mass loaded wind profiles, calculated using $\lambda_{\rm ML}= 5 d_0$ (thin lines) and $\lambda_{\rm ML}= 30 d_0$ (thick lines), and the profile of the unperturbed bow shock, calculated using the thin shock approximation (dot-dashed line) \citep{Wilkin96}.  The shaded region, with radius $r=d_0$, shows the unperturbed pulsar wind, i.e. without the mass loading effect.}
\label{fig2}
\end{figure}

The behavior of $A$ and $u$ as a function of the normalized distance is shown in the left panel of Fig.\ref{fig2}, where we chose $V_0 = 300$ km s$^{-1}$, $u_1=c$ and $n_0= 0.1$ cm$^{-3}$. As one would expect, the loaded mass slows down the wind on a typical length scale of $\lambda_{\rm ML}$ and, because of Eq.(\ref{eq:A(u)}), the cross section expands. For distances $x \gg \lambda_{\rm ML}$ we have $u \rightarrow V_0$ and $A/A_1 \rightarrow u_1/V_0$. Because $u_1\approx c \gg V_0$, the wind cross section can expand up to a factor $\approx 10^3$.

In the right panel of Fig.(\ref{fig2}) we compare our solutions for the wind profile with the unperturbed wind and with the unperturbed bow shock profile as predicted by \citep{Wilkin96}. For $\lambda_{\rm ML} \lesssim 100 d_0$ the modified wind crosses the bow shock, implying that the bow shock shape will be strongly affected.

The left panel of Fig.\ref{fig2} also reports the velocity of the expansion in the radial direction, derived from the evolution of $A$ and $u$ with $r$, i.e. $u_r \equiv \frac{dr}{dt} = \frac{V_0}{2 \sqrt{\pi A}} \frac{dA}{du} \frac{du}{dx}$. It is easy to show that the value of $u_r$ at the origin is $u_{r1}  = V_0 d_0/(2 \lambda_{\rm ML})$, while for very large distances one has $u_r \rightarrow 0$. Moreover the radial expansion speed has a peak at the position $x_{\rm peak}= \lambda_{\rm ML} \left(0.25+\ln[u_1/(4V_0)] \right)$ where its value is $u_{r, \rm peak}=0.3\, u_{r1} \sqrt{u_1/V_0}$ . This behavior has two important consequences. The first one is that for $\lambda_{\rm ML} \lesssim 100 d_0$ the expansion speed is larger than the sound speed in the ISM when the wind crosses the unperturbed bow shock profile. In other words, when the pulsar wind expands beyond the unperturbed bow shock, the expansion is supersonic and one expects a strong modification of the bow shock. By contrast, for $\lambda_{\rm ML} \gtrsim 100 d_0$ the expansion of the wind in the ISM is subsonic, and one therefore expects a less pronounced deformation of the bow shock profile. 
The second consequence is that for distances close to $x_{\rm peak}$, $u_r$ can be even larger than $u$.
When the latter condition is realized the quasi 1-D approximation is no longer valid and our model can no longer be used.
Based on the above arguments one may speculate that the system becomes non-stationarity at that location, giving rise to a periodic structure of expanding bubbles, similar to those observed in the Guitar Nebula.  However, addressing this scenario using only analytical models is a complicated matter, requiring the use of full 2-D numerical simulations.

\end{document}